\documentclass[%
superscriptaddress,
reprint,
amsmath,amssymb,
aps,
pra,
]{revtex4-2}

\usepackage[english]{babel}

\usepackage{graphicx}
\usepackage[percent]{overpic}
\usepackage{dcolumn}
\usepackage{bm}
\usepackage{bbm}
\usepackage{amsmath}
\usepackage{amssymb}
\usepackage{mathtools}
\usepackage{physics}
\usepackage{siunitx}
\usepackage[version=4]{mhchem}
\usepackage{dsfont}
\usepackage{upgreek}

\usepackage{import}

\usepackage{hyperref}
\usepackage{color}
\definecolor{rltred}{rgb}{0.75,0,0}
\definecolor{rltgreen}{rgb}{0,0.6,0}
\definecolor{rltblue}{rgb}{0.3,0.3,1}

\hypersetup{colorlinks,%
	hypertexnames=true,%
	linkcolor=rltred,%
	citecolor=rltgreen,%
	linktocpage=true}

\newcommand{\reals}{\mathbb{R}}

\newcommand{\vectbf}[1]{{\mathbf{#1}}}

\DeclareMathOperator{\diag}{diag}

\DeclareSIUnit\bar{bar}

\begin{document}
\title{Coupling free electrons to a trapped-ion quantum computer}

\author{Elias Pescoller}
\email{elias.pescoller@tuwien.ac.at}
\affiliation{Institute for Theoretical Physics, TU Wien, Wiedner Hauptstra\ss e 8-10/136, 1040 Vienna, Austria}
\affiliation{Vienna Center for Quantum Science and Technology, Atominstitut, TU Wien, Stadionallee 2, 1020 Vienna, Austria}
 
\author{Santiago Beltr\'an-Romero}
\affiliation{Vienna Center for Quantum Science and Technology, Atominstitut, TU Wien, Stadionallee 2, 1020 Vienna, Austria}
\affiliation{University Service Centre for Transmission Electron Microscopy, TU Wien, Stadionallee 2, 1020 Vienna, Austria}

\author{Sebastian Egginger}
\affiliation{Johannes Kepler University Linz, 4040 Linz, Austria}

\author{Nicolas Jungwirth}
\affiliation{University of Innsbruck, 6020 Innsbruck, Austria}

\author{Martino Zanetti}
\affiliation{University of Vienna, Faculty of Physics, Vienna Center for Quantum Science and Technology, 1090 Vienna, Austria}
\affiliation{University of Vienna, Max Perutz Labs, 1030 Vienna, Austria}
\affiliation{Vienna Center for Quantum Science and Technology, Atominstitut, TU Wien, Stadionallee 2, 1020 Vienna, Austria}

\author{Dominik Hornof}
\affiliation{Vienna Center for Quantum Science and Technology, Atominstitut, TU Wien, Stadionallee 2, 1020 Vienna, Austria}
\affiliation{University Service Centre for Transmission Electron Microscopy, TU Wien, Stadionallee 2, 1020 Vienna, Austria}

\author{Michael S. Seifner}
\affiliation{Vienna Center for Quantum Science and Technology, Atominstitut, TU Wien, Stadionallee 2, 1020 Vienna, Austria}
\affiliation{University Service Centre for Transmission Electron Microscopy, TU Wien, Stadionallee 2, 1020 Vienna, Austria}

\author{Iva B\v rezinov\'a}
\email{iva.brezinova@tuwien.ac.at}
\affiliation{Institute for Theoretical Physics, TU Wien, Wiedner Hauptstra\ss e 8-10/136, 1040 Vienna, Austria}

\author{Philipp Haslinger}
\email{philipp.haslinger@tuwien.ac.at}
\affiliation{Vienna Center for Quantum Science and Technology, Atominstitut, TU Wien, Stadionallee 2, 1020 Vienna, Austria}
\affiliation{University Service Centre for Transmission Electron Microscopy, TU Wien, Stadionallee 2, 1020 Vienna, Austria}
    
\author{Thomas Juffmann}
\email{thomas.juffmann@univie.ac.at}
\affiliation{University of Vienna, Faculty of Physics, Vienna Center for Quantum Science and Technology, 1090 Vienna, Austria}
\affiliation{University of Vienna, Max Perutz Labs, 1030 Vienna, Austria}

\author{Johannes Kofler}
\email{johannes.kofler@jku.at}
\affiliation{Johannes Kepler University Linz, 4040 Linz, Austria}

\author{Philipp Schindler}
\email{philipp.schindler@uibk.ac.at}
\affiliation{University of Innsbruck, 6020 Innsbruck, Austria}

\author{Dennis R\"atzel}
\email{dennis.raetzel@tuwien.ac.at}
\affiliation{Vienna Center for Quantum Science and Technology, Atominstitut, TU Wien, Stadionallee 2, 1020 Vienna, Austria}
\affiliation{University Service Centre for Transmission Electron Microscopy, TU Wien, Stadionallee 2, 1020 Vienna, Austria}
\affiliation{ZARM, University of Bremen, Am Fallturm 2, 28359 Bremen, Germany}
    
\date{\today}

\begin{abstract}
Freely propagating electrons may serve as
quantum probes that can
become coherently correlated with other quantum systems, offering access to advanced metrological resources.
We propose a setup that coherently 
couples free electrons
in an electron microscope to a trapped-ion quantum processor, enabling non-destructive, quantum-coherent
detection and the accumulation of information across multiple electrons. Our analysis shows that
single electrons can induce resolvable qubit excitations,
establishing a
platform for practical applications such as quantum-enhanced, dose-efficient electron microscopy.
\end{abstract}

\maketitle

Electron microscopy (EM)~\cite{ReimerKohl2008} has reached extraordinary spatial~\cite{ishikawa2023spatial,suenaga_element-selective_2000} and temporal~\cite{zewail2000femtochemistry,baum2007attosecond,kealhofer2016all,FEIST201763,gaida2024attosecond} resolution, enabling direct visualization of matter at the atomic scale. These advances have been driven by major developments in electron optics, including aberration correction and high-brightness sources~\cite{ishikawa2023spatial, muller_atomic-scale_2008},
revolutions in electron detection~\cite{fan2000digital, mcmullan_direct_2016, llopart2022timepix4_electron_detection, ishikawa2023spatial}, and increasingly sophisticated computational reconstruction techniques~\cite{kirkland1998computing}. As a result, EM has become a powerful, high-throughput imaging method in both biology~\cite{kuhlbrandt_biochemistry_2014, glaeser_how_2016} and materials science~\cite{chen_electron_2021}. 
In the near future, further advances may be provided, for example, by laser-based electron wave packet shaping which has already been experimentally demonstrated \cite{Schwartz2019, Mihaila2022, Mihaila2025}.

Despite this progress, the information that can be extracted from a specimen remains fundamentally constrained by the interaction between the electron beam and the sample. Inelastic scattering processes induce structural and chemical modifications, imposing a finite tolerable electron dose~\cite{henderson_potential_1995,glaeser_limitations_1971,egerton_radiation_2004,isaacson_electron_1973} and, consequently, an upper bound on the achievable signal-to-noise ratio~\cite{bouchet_fundamental_2021, dwyer_quantum_2023}. This limitation is particularly severe for radiation-sensitive specimens such as living cells, vitrified biological material, viruses, and individual proteins. Improving EM under such dose constraints therefore requires strategies that increase the information gained per electron, rather than simply increasing beam intensity.

\begin{figure*}
    \includegraphics[width=\linewidth]{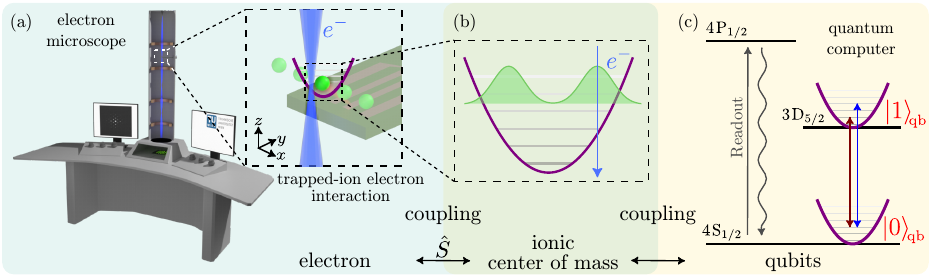}
    \caption{(a) A transmission electron microscope (TEM). Within a plane conjugate to the specimen plane, a trapped-ion quantum computer is inserted. (b) The electron $e^-$ couples to the ionic center-of-mass degree of freedom via the Coulomb interaction (scattering matrix $\hat{S}$). If the ion is prepared in a superposition of coherent states, the electron induces a relative phase shift between these states. (c) The figure shows a simplified level diagram of \ce{^40Ca+} with the relevant transitions used to encode the qubit. The center-of-mass degree of freedom of the ion is coupled to the 4S$_{1/2}$ and 3D$_{5/2}$ states of the valence electron (qubit degree of freedom) via the application of bichromatic laser pulses (red- and blue-detuned transitions in the figure). The rapidly decaying fluorescent transition from 4P$_{1/2}$ to 4S$_{1/2}$ is used to perform projective readout of the qubit state.}
    \label{fig:sketches}
\end{figure*}
At the same time, electron microscopy has largely operated within a semiclassical measurement paradigm, in which quantum correlations involving the probing electrons are neglected or treated as detrimental. Yet, the entanglement of electrons with other quantum systems, including electromagnetic fields, detectors, and, in principle, the specimen itself, may be harnessed to open up an additional and largely unexplored avenue for enhancing metrological performance, enabling new measurement modalities and protocols that go beyond classical bounds \cite{mechel2021quantum,zhao_quantum_2021,ruimy2025free,abajo_roadmap_2025,synanidis_rydberg-atom_2025,Nimmrichter2025elec,even2025spin,maison2025cavity,Rembold_2025}.

Recent theoretical proposals and experimental demonstrations illustrate how both challenges, i.e., low-dose operation and quantum-enhanced measurements can be addressed: Approaches such as employing entangled multi-electron probe states~\cite{koppell_transmission_2022}, coherent multi-pass electron microscopy~\cite{juffmann_multi-pass_2017}, interaction-free measurements~\cite{elitzur_quantum_1993, putnam_noninvasive_2009, kruit_designs_2016, turner_interaction-free_2021}, entanglement-assisted phase contrast~\cite{okamoto_entanglement-assisted_2014}, and quantum search-inspired measurement protocols~\cite{okamoto_universal_2023} aim to maximize information extraction per damaging interaction. Complementarily, coherent interactions of free electrons with other quantum systems such as plasmons \cite{vesseur2007direct}, spins ~\cite{jarovs2025SpinSensing, haslinger2024spin,ratzel_controlling_2021, kolb2025QUAK}, and photons \cite{barwick2009photon,feist2015quantum,Wang2020,dahan2021imprinting,feist2022cavity,Konecna2022} including electron-photon entanglement ~\cite{henke2025observationquantumentanglementfree, preimesberger_experimental_2025} 
have been experimentally demonstrated, highlighting the feasibility of coherent electron–quantum-system interfaces. Together, these developments point toward a new regime of quantum-enhanced electron microscopy~\cite{abajo_roadmap_2025}, in which dose efficiency and quantum metrology are intrinsically linked.

Motivated by the rapid progress of quantum information processors, we propose a concrete implementation of this paradigm by coherently coupling free electrons in an electron microscope to a trapped-ion quantum computer. We analyze the interaction between electrons and trapped-ions and propose a protocol to entangle free electrons with trapped-ion qubits.
Beyond enabling the preparation of advantageous quantum probe states, this interface grants access to the full space of quantum measurements through the universal state-manipulation capabilities of the quantum processor, establishing a versatile platform for low-dose, quantum-enhanced electron microscopy. 
The proposed platform can also be useful for coherent state preparation and readout in other setups that employ free electrons, for example, quantum free electron lasers \cite{kling2015defines,BONIFACIO200869} and nano-scale electron accelerators \cite{shiloh2022miniature}. Furthermore, the coupling scheme may be generalized to different highly controlled charged particles ranging from focused ion beams to the protons used in high energy physics.

In the following, we assume a focused electron wavefunction, as commonly produced and coherently shaped within a transmission electron microscope (TEM), see Fig.~\ref{fig:sketches} (a), interacting with ions trapped in a harmonic oscillator potential. 
A promising ion-trap platform consists of a planar surface-electrode Paul trap on top of an integrated photonic chip, which both confines \ce{^{40}Ca+} ions around hundred micrometers above the surface and manipulates them with lasers through grating couplers [see Fig.~\ref{fig:sketches} (a), more details on integrating an ion trap experiment into a TEM are given in App.~\ref{ap:tem_iontrap}]. This choice has the advantage that the ions can be separated by several hundred micrometers from each other~\cite{mosesRaceTrack20223}, allowing for the electron to couple to single ions in a larger quantum register. The ion trap includes a sufficiently large opening around the trapping region, allowing the electron beam to pass by.
Typical trap frequencies $\Omega$ lie between $0.5~\mathrm{MHz}$ and $5~\mathrm{MHz}$ which correspond to harmonic-oscillator ground-state sizes 
$R_0/\sqrt{2}$ of $40~\mathrm{nm}$ and $13~\mathrm{nm}$, respectively, where
 $R_0=\sqrt{\hbar/m_\text{ion}\Omega}$. After resolved-sideband cooling, the ion can be initialized in its motional and electronic ground state \cite{schindler_quantum_2013}.

The \ce{^40Ca+} ions provide an established platform for quantum information processing, including coherent manipulation on the narrow S-D quadrupole transition and projective readout on the S-P cycling transition [Fig.~\ref{fig:sketches} (c)]. These quantum operations have been demonstrated in compact experimental systems with more than 15 ions \cite{Pogorelov2021}. 
Operations between multiple qubits are performed by bringing the ions close together and coupling their qubit states to their common motion in a single trap potential. The same tools can be used to create non-classical sensing states involving the qubit of a single ion and its motion \cite{hempel_entanglement-enhanced_2013}.

The coupling between the electrons and the qubits is then mediated by the ionic center of mass motion, which couples directly to the free electron via the Coulomb interaction and to the qubits via the application of laser pulses, see Fig.~\ref{fig:sketches}. The focus of the present paper is the analysis of the first stage of this coupling scheme described in the following by the scattering operator $\hat S$.

We first model the interaction of a single electron with a single trapped ion.
We start from the Hamiltonian
$
    \hat H = \hat H_\text{el} + \hat H_\text{cmi} + \hat H_\text{qb} + \hat H_\text{int},
$
where 
$
    \hat H_\text{el} = c\sqrt{c^2+\vectbf{\hat p}^2}
$
is the relativistic energy of the free electron (in atomic units, $\hbar=m_\text{e}=\abs{e}=4\pi\epsilon_0=1$),
$\vectbf{\hat p}$ is the electron momentum operator, 
$
    \hat H_\text{cmi}
$
is the kinetic and potential energy of the center-of-mass degree of freedom of a harmonically trapped ion with trap frequencies $\Omega_x=\Omega_y=\Omega$ and $\Omega_z$, and $H_\text{qb}=\omega\hat\sigma^z_\text{qb}$ is the free Hamiltonian for the two-level system used to encode the qubit.
The operator
$
    \hat H_\text{int} =
    -1/\abs*{\vectbf{\hat r}-\vectbf{\hat R}}
$
is the Coulomb interaction of the free electron ($\vectbf{\hat r}$) with the ionic center-of-mass ($\vectbf{\hat R}$). 
Note that the probability for internal transitions of the ion induced by the free electron is negligibly small in the relevant parameter regime (see App.~\ref{ap:decoherence_internal}). Therefore, we do not include the direct interaction of the electron with the internal degrees of freedom of the ion into the model.

In a TEM, the electron momentum is typically dominated by a large component in the direction of $-\vectbf{e_z}$, where we fix the coordinate system such that the $z$-axis is aligned with the beam axis [Fig.~\ref{fig:sketches} (a)]. To leading order in the relative transverse and longitudinal momentum spreads of the electron wave packet, this allows us to replace the electron momentum operator $\vectbf{\hat p}$ in the interaction Hamiltonian by its average $\vectbf{p_\text{el}}=-\gamma v_\text{el}\vectbf{e_z}$ corresponding to the paraxial approximation. Furthermore, the interaction time with the ion is much shorter than the timescale of the ionic center-of-mass motion, such that we can use the stroboscopic approximation, i.e.~neglecting the time-evolution of the ion during the scattering event, to derive the scattering operator [see App.~\ref{ap:scattering_operator}]
\begin{equation}\label{eq:scattering_operator}
    \hat S = \exp(-\frac{2i}{v_\text{el}}\log\abs*{\vectbf{\hat r_\perp}- \vectbf{\hat R}^{\text{(I)}}_\perp(t_\text{el})}).
\end{equation}

Here, $\vectbf{\hat r_\perp}$ is the transverse position
operator for the electron, $\vectbf{\hat R}_\perp^{\text{(I)}}(t_\text{el})=\vectbf{\hat R}_\perp\cos(\Omega t_\text{el})+\vectbf{\hat P}_\perp/m_\text{ion}\sin(\Omega t_\text{el})$ is the transverse position
operator of the ion in the interaction picture, $\vectbf{\hat R}_\perp$ and $\vectbf{\hat P}_\perp$ are the transverse position
and momentum operators of the ion, respectively, and
$t_\text{el}$ is the electron arrival time.

With the approximations made (paraxial and stroboscopic), the scattering operator only couples the transverse degrees of freedom of the electron and the ion. For the subsequent analysis, we will thus express the joint wavefunction of the electron, the center-of-mass motion of the ion, and the qubits in terms of states of the form
$
\ket{\psi_\vectbf{r_\perp}}_\text{el}\ket{\boldsymbol{\upalpha}}_\text{cmi}\ket{\sigma}_\text{qb}.
$
Here, $\ket{\psi_\vectbf{r_\perp}}_\text{el}$ denotes a focused Gaussian electron state in the plane of the ion trap, with midpoint $\vectbf{r_\perp}=(x,y)$ and width $\delta_{\vectbf{r}_\perp}$. We assume the width $\delta_{\vectbf{r}_\perp}$ to be significantly smaller than the zero-point fluctuation of the trapped ion $\delta_{\vectbf{R}_\perp} = R_0/\sqrt{2} = 1/\sqrt{2m_\text{ion}\Omega}$. The ion center-of-mass state $\ket{\boldsymbol{\upalpha}}_\text{cmi}$ corresponds to a coherent state in the harmonic trap with displacements $\boldsymbol{\upalpha}=(\alpha_x,\alpha_y)$. Finally, $\ket{\sigma}_\text{qb} \in\{\ket{0}_\text{qb},\ket{1}_\text{qb}\}$ is the qubit encoded in the internal electronic degrees of freedom of the ion.

We now study the effect of this operator on the basis states of the form 
$
\ket{\psi_\vectbf{r_\perp}}_\text{el}\ket{\boldsymbol{\upalpha}}_\text{cmi}\ket{\sigma}_\text{qb}.
$
As long as the spread of the electronic wavefunction is small enough ($\delta_{\vectbf{r}_\perp} \lesssim 0.1~\delta_{\vectbf{R}_\perp} \lesssim 1\,\si{\nano\meter}$), the focused electronic states $\ket{\psi_\vectbf{r_\perp}}_\text{el}$ are approximate eigenstates of the scattering operator. In this setting, we may ignore any back action, including momentum kicks on the electron. In App.~\ref{ap:backaction} we discuss the effect of a finite extent of the electronic wavefunction and give upper bounds to the decoherence arising from the back action to the electron.

\begin{figure}
    \includegraphics[width=\linewidth]{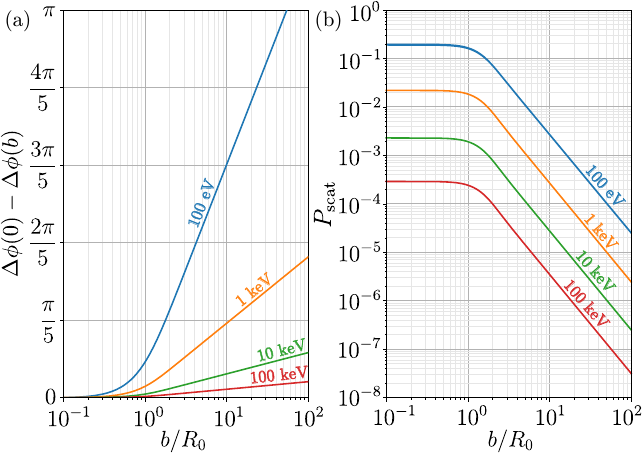}
    \caption{
    (a) The phase shift $\Delta\phi(b)$ (Eq.~\eqref{eq:action_on_coherent}), assuming $\delta_{\vectbf{r_\perp}} \ll R_0$, of the combined electron-ion wavefunction due to the interaction with impact parameter $b$ compared to the corresponding phase shift $\Delta\phi(0)$ when the electron is focused to the center of the harmonic trap.
    (b) The probability 
    $
        P_\text{scat}=1-\abs*{
            \mathcal{S}(\vectbf{r}_\perp,\boldsymbol{\upalpha})
        }^2
    $, assuming $\delta_{\vectbf{r_\perp}} \ll R_0$,
    for the ion to be scattered out of the state $\ket{\boldsymbol{\upalpha}}_\text{cmi}$ upon impact of an electron as a function of the impact parameter $b$.}
    \label{fig:interaction_plot}
\end{figure}
The main effect of the scattering operator Eq.~\eqref{eq:scattering_operator} acting on coherent states is to imprint a displacement-dependent phase shift onto them:
\begin{equation}
\label{eq:action_on_coherent}
    \hat S\ket{\psi_\vectbf{r_\perp}}_\text{el}\ket{\boldsymbol{\upalpha}}_\text{cmi}
    \approx e^{i\Delta\phi(b)}\ket{\psi_\vectbf{r_\perp}}_\text{el}\ket{\boldsymbol{\upalpha}}_\text{cmi},
\end{equation}
with the impact parameter $b=\abs*{\vectbf{r_\perp}-\sqrt{2}R_0\Re(\boldsymbol{\upalpha}e^{i\Omega t_\text{el}})}$.
The phase shift
$
    \Delta\phi=\arg(\mathcal{S}(\vectbf{r}_\perp,\boldsymbol{\upalpha}))
$
is given by the complex argument of the state overlap of initial and final state:
$
\mathcal{S}(\vectbf{r}_\perp,\boldsymbol{\upalpha}) =
    \bra{\psi_\vectbf{r_\perp}}_\text{el}\bra{\boldsymbol{\upalpha}}_\text{cmi}\hat S\ket{\boldsymbol{\upalpha}}_\text{cmi} \ket{\psi_\vectbf{r_\perp}}_\text{el}
$.
In Fig.~\ref{fig:interaction_plot} (a) we plot $\Delta\phi$ as a function of the impact parameter and assuming $\delta_{\vectbf{r_\perp}} \ll R_0$ and we also give an analytical expression for $\Delta\phi(b)$ in App.~\ref{ap:explicit_computation_of_phase}.
There is also a finite probability for the scattering process to change the state of the ion by more than just a phase,
given by
$
P_\text{scat}=1-\abs*{
\mathcal{S}(\vectbf{r}_\perp,\boldsymbol{\upalpha})
}^2
$ and plotted, again assuming $\delta_{\vectbf{r_\perp}} \ll R_0$, in Fig.~\ref{fig:interaction_plot} (b).
The probability for this to happen is approximately constant for impact parameters $b<R_0$, with $P_\text{scat}=1-\pi/(v_\text{el}\sinh(\pi/v_\text{el}))$ for when the electron is focused to the center of the trap ($b=0$) [See App.~\ref{ap:explicit_computation_of_phase}], and decays quadratically for larger impact parameters.
For $b=0$ and an electron energy of $100~\si{\eV}$ the probability becomes approximately $10^{-1}$, decreasing rapidly for higher energies.
In the following, we focus on the induced phase shift and assume the probability of state changes to be negligible.

The phase imprinted onto the joint electron-ion wavefunction, Eq.~\eqref{eq:action_on_coherent}, is a global phase that is undetectable. However, one may prepare the ion in a qubit-motion entangled cat state, that is, a superposition of two coherent states with displacements $-\boldsymbol{\upalpha}$ and $\boldsymbol{\upalpha}$, entangled with the qubit degree of freedom: $\ket{\psi_\text{cat}} = (\ket{-\boldsymbol{\upalpha}}_\text{cmi}\ket{-}_\text{qb}+\ket{\boldsymbol{\upalpha}}_\text{cmi}\ket{+}_\text{qb})/\sqrt{2}$. Here, $\ket{\pm}_\text{qb}=(\ket{0}_\text{qb}\pm\ket{1}_\text{qb})/\sqrt{2}$ are the eigenstates of the Pauli-X operator $\hat\sigma^x_\text{qb}$. 
It has been experimentally shown that such a state can be prepared by using bichromatic laser pulses [see Fig.~\ref{fig:sketches} (c)], resulting in a qubit-dependent displacement \cite{hempel_entanglement-enhanced_2013}. This operation can be represented as:
\begin{equation}
\label{eq:qubit_dependent_displacement}
    \hat D(\hat \sigma^x_\text{qb}\boldsymbol{\upalpha})=\exp(\hat \sigma^x_\text{qb} \left(\boldsymbol{\upalpha}\cdot \vectbf{\hat a}^\dag - \boldsymbol{\upalpha}^\ast \cdot \vectbf{\hat a}\right)),
\end{equation}
where
$\vectbf{a}$ and $\vectbf{a}^\dag$ are the (transverse) ladder operators in the harmonic trap. 
We assume the electron arrival to be synchronized with the harmonic trap such that at time $t_\text{el}$, the ion is at its turning point: $\boldsymbol{\upalpha}e^{i\Omega t_\text{el}}\in\reals^2$. For the sake of readability, we also fix $t_\text{el}=0$, such that $\boldsymbol{\upalpha}$ is real.
Then, the final ionic wavefunction after tracing out the electronic degrees of freedom becomes (up to a global phase): 
\begin{equation}
\label{eq:final_cat_state}
    \ket{\psi_\text{cat}'} = \frac{\ket{-\boldsymbol{\upalpha}}_\text{cmi}\ket{-}_\text{qb}+e^{ig(\vectbf{r}_\perp, {\boldsymbol{\upalpha}})}\ket{\boldsymbol{\upalpha}}_\text{cmi}\ket{+}_\text{qb}}{\sqrt{2}}.
\end{equation}
Here, $g(\vectbf{r}_\perp, {\boldsymbol{\upalpha}})\!=\!\Delta\phi(\abs*{\vectbf{r}_\perp\!-\!\sqrt{2}R_0\boldsymbol{\upalpha}})\!-\!\Delta\phi(\abs*{\vectbf{r}_\perp\!+\!\sqrt{2}R_0\boldsymbol{\upalpha}})$ is the relative phase shift due to the difference in impact parameter.
Thus, the final ionic state is distinguishable from the initial state, with $\arccos(\abs*{\braket{\psi_\text{cat}}{\psi_\text{cat}'}})=g(\vectbf{r}_\perp, {\boldsymbol{\upalpha}})/2$. The quantum computer (in form of the ion) has thus acquired knowledge about the presence of an electron.
The maximal coupling strength is reached when the electron beam is focused to the center of one of the two coherent states [See Fig.~\ref{fig:sketches} (b)]. For $\vectbf{r}_\perp = \sqrt{2}R_0\boldsymbol{\upalpha}$ we get
$
    g(\sqrt{2}R_0\boldsymbol{\upalpha},\boldsymbol{\upalpha})=\Delta\phi(0)-\Delta\phi(2\sqrt{2}R_0\abs*{\boldsymbol{\upalpha}})
$.
\begin{figure}
    \includegraphics[width=\linewidth]{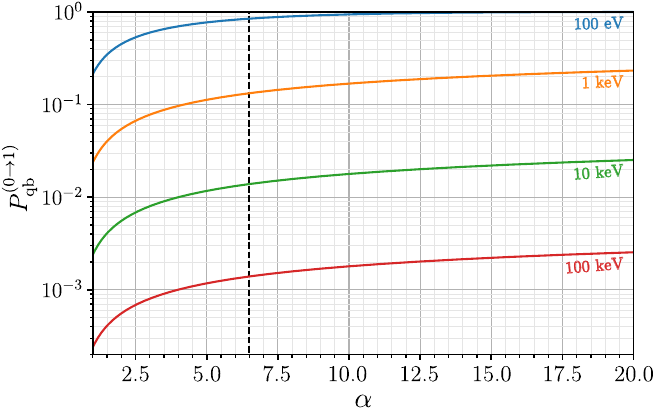}
    \caption{The plot shows the probabilities to find the qubit in the state $\ket{1}$ after the interaction with an electron of different energies as a function of the cat state size $\alpha$. It is assumed that the qubit is initially prepared in the state $\ket{0}$. The dashed line indicates $\abs*{\boldsymbol{\upalpha}}=6.5$, which has already been demonstrated experimentally \cite{wu2025infraredabsorptionspectroscopysingle}.
    }
    \label{fig:qubit_flip_probability}
\end{figure}

After the interaction with the electron, the two coherent states can be recombined by inverting the operation
Eq.~\eqref{eq:qubit_dependent_displacement}. Altogether, assuming the ion's center of mass to initially be in the ground state, this results in an effective unitary $\hat U_\text{e--qb} = \bra{0}_\text{cmi}{\hat D(-\hat \sigma^x_\text{qb}\boldsymbol{\upalpha})\hat S\hat D(\hat \sigma^x_\text{qb}\boldsymbol{\upalpha})}\ket{0}_\text{cmi}$ that couples the qubit to the electron by performing a rotation on the qubit that depends on the position of the electron $\vectbf{r_\perp}$:
\begin{equation}
    \label{eq:qubit_scattering_operator}
    \hat U_\text{e--qb} \approx 
    e^{i\kappa(\vectbf{\hat r_\perp},\boldsymbol{\upalpha})}e^{\frac{i}{2} \hat\sigma^x_\text{qb}
    g(\vectbf{\hat r_\perp},\boldsymbol{\upalpha})}.
\end{equation}
The qubit-independent phase $\kappa$ is thereby given by
$
\kappa(\vectbf{r}_\perp, {\boldsymbol{\upalpha}})\!=\!(\Delta\phi(\abs*{\vectbf{r}_\perp\!-\!\sqrt{2}R_0\boldsymbol{\upalpha}})\!+\!\Delta\phi(\abs*{\vectbf{r}_\perp\!+\!\sqrt{2}R_0\boldsymbol{\upalpha}}))/2
$.
By implementing this protocol, it is thus possible to produce an effective coupling between the free probe electron and the bound electronic qubit. 
The strength of the coupling can be understood in terms of the probability to measure the qubit in the $\ket{1}_\text{qb}$-state after the interaction, assuming that it was initially prepared in the $\ket{0}_\text{qb}$-state. We plot this probability in Fig.~\ref{fig:qubit_flip_probability}. For slow electrons ($100~\si{\eV}$ -- $1~\si{\kilo\eV}$) and experimentally plausible values for $\abs*{\boldsymbol{\upalpha}}$, the probability for a bit-flip reaches the order of $0.1$ -- $1$.
While not usually employed in TEM, these electron energies are well matched to the technological advances of the last decades in electron microscopy.
Atomic resolution has been demonstrated down to approximately $15~\si{\kilo\eV}$~\cite{SALVE2016,15keV_atomic_2016} and even lower electron energies, down to effectively $0~\si{\eV}$, are utilized in high-resolution EMs, achieving spatial resolutions down to $2~\si{\nano\meter}$ via beam acceleration and deceleration stages \cite{tromp2010new,LEEM_eVTEM2021}. Spot sizes down to $2~\si{\nano\meter}$ can be reached in scanning electron microscopy \cite{FRANK201146}, showing that it is possible to focus the electron to the center of the ionic wavefunction. Thus, by combining the setup with additional acceleration and deceleration stages it would be possible to reach the strongly interacting regime without sacrificing the high resolution. Furthermore, in multi-pass geometries \cite{juffmann_multi-pass_2017}, the effect due to multiple interactions can add up coherently, rendering even $100~\si{\kilo\eV}$ an attractive parameter regime.

For the interaction with multiple electrons, Eq.~\eqref{eq:qubit_scattering_operator} readily generalizes to the product
\begin{equation}
    \hat U_{N_{\rm el}-\text{qb}} \approx e^{i\sum_{k=1}^{N_\text{el}}\kappa(\vectbf{\hat r}_\perp^\text{(k)}, \boldsymbol{\upalpha})}
   e^{\frac{i}{2}\hat \sigma_\text{qb}^x\sum_{k=1}^{N_\text{el}}g(\vectbf{\hat r}_\perp^\text{(k)}, \boldsymbol{\upalpha})}.
\end{equation}
This establishes the trapped ion as a quantum detector that is able to interact with multiple electrons in a coherent way. Such a detector may be used to generate many-electron entanglement as well as exotic ionic and electronic states in a two-path setup \cite{ruimy_many-body_2024}.

So far we have only discussed the interaction with a single ion. On a quantum computer based on surface-ion traps, one typically has many ions that can be arranged in linear or even quadratic arrays \cite{holz_2d_2020}. In a linear arrangement of the ions aligned with the beam axis, the coherent interaction with the ion chain results in a collective enhancement of the coupling strength, scaling with the number of ions. Arranging the ions perpendicular to the beam axis as visualized in Fig.~\ref{fig:sketches}(a), on the other hand, can give rise to many new measurement schemes.
We envision a setup in which the electron is prepared in a superposition of distinct paths, where each of the paths is focused onto a single ion. This could be achieved by using electron diffraction gratings, as proposed in \cite{kruit_designs_2016}.
In each path, the electron would then interact with the corresponding ion according to Eq.~\eqref{eq:qubit_scattering_operator}, yielding the one-electron-many-qubit 
coupling operator:
\begin{equation}
    \label{eq:many_qubit_scattering_operator}
    \hat U_{e-N_{\rm qb}}\approx 
    e^{i\kappa} \sum_{k=1}^{N_{\rm qb}}
    e^{\frac{i}{2}g\hat \sigma^x_{\text{qb}, k}}\ketbra{\psi_{k}}_\text{el}.
\end{equation}
Here, $\ket{\psi_{k}}$ denotes the $k^\text{th}$ electron path and $\hat\sigma^x_{\text{qb}, k}$ acts on the $k^\text{th}$ qubit. The coupling strength $g=\Delta\phi(0)-\Delta\phi(2\sqrt{2} R_0\abs*{\boldsymbol{\upalpha}})$ as well as the global phase $\kappa$ is constant here as we assume the individual beams to be focused to the center of the corresponding coherent state $\ket{\boldsymbol{\upalpha}_k}_\text{cmi}$ and we fix the magnitude of $\boldsymbol{\upalpha}_k$.
The operator Eq.~\eqref{eq:many_qubit_scattering_operator} can be understood as follows: If the electron is in the $k^{\text{th}}$ path, the $k^{\text{th}}$ qubit is subjected to a partial bit-flip. 
By these means, the quantum computer's state becomes entangled with the electron path degree of freedom.
If the electron subsequently traverses a weak-phase-object, i.e.\ a biological specimen, the joint wavefunction acquires a slight phase shift.
The electron-ion entanglement can now be used to transfer this phase shift to the qubits. This can be achieved by measuring the electron in momentum space and performing appropriate measurement-result dependent operations on the quantum computer, see App.~\ref{ap:phase_kick_back}.

Acquiring phase information about the specimen in this way might have many benefits. Since the coupling is coherent, phase shifts due to multiple probe electrons can add up \cite{okamoto_entanglement-assisted_2014}, making it possible to surpass the standard quantum limit. In App.~\ref{ap:example1} we explain briefly how the scheme proposed in \cite{okamoto_entanglement-assisted_2014} can be adapted to the ion-trap platform and discuss how to generalize it to allow for electron losses and non-ideal coupling strengths. Furthermore, by coupling the electron to the trapped-ion-array one can perform arbitrary quantum operations to extract as much information as possible out of a single electron. 
A similar advantage has been claimed in \cite{okamoto_quantum_2006}, where specially engineered diffractive elements are proposed as a means to perform optimal measurements, and in \cite{okamoto_universal_2023}, where it has been proposed to implement Grover-like detection algorithms.

If combined with a multi-pass electron-microscopy setup, the quantum computer might provide even more advantages. When the electron interacts only once with the ion and is then measured by the classical detector of the electron microscope, coherence can only be assured if the measurement result is used to compensate for the phase imprinted on the ion by the interaction.
In a multi-pass setup, this requirement would drop out as the electron can interact many times with both sample and quantum computer without being detected. Furthermore, in the multi-pass electron microscopical setups currently envisioned \cite{juffmann_multi-pass_2017, KOPPELL2019112834}, the probe electron has to be coupled out of the resonator in order to extract information. By using a trapped-ion quantum computer as a detector, a continuous coherent information transfer from the specimen to the quantum computer, mediated by a single or few electrons, could be possible.
This would allow for weak or strong measurements without coupling the electron out of the resonator.

Our approach thus represents a significant step towards an integration of a well-established quantum-computing platform \cite{cirac_quantum_1995} into the powerful and versatile framework of electron microscopy.

\vspace{1em}\section*{Acknowledgements}
We thank the members of USTEM and Richard Kueng for useful discussions. This research is funded in part by the Gordon and Betty Moore Foundation Grant GBMF12992, 
by the Austrian Science Fund (FWF) through Grant No.~10.55776/COE1, Y1121, P36041, P35953 and the FFG-project AQUTEM.

\bibliography{qcem}

\clearpage

\appendix

\section{Integration of an ion trap into a TEM}
\label{ap:tem_iontrap}

In this appendix, we discuss several experimental challenges that arise from integrating an ion trap into a TEM. 

First of all, to operate the ion trap inside the TEM, it is necessary to reduce the pressure at the trap to $\sim\!10^{-9}~\si{\milli\bar}$ or better. Lower pressures result in longer trapping and coherence times of the ion quantum state. Similar vacuum conditions have already been achieved in modified electron microscopes~\cite{LEUTHNER201976, heinmann1986, MISHIMA1998L256}, typically relying on all-metal vacuum chambers and increased pumping speeds. In our case, the pumping speed can be increased by adding combined ion getter (IG) and non-evaporable getter (NEG) pumps in the vicinity of the sample area, as well as through cryopumping. NEG pumps are particularly effective as they feature very high pumping speed for species like \ce{H_2} compared to the IG pump originally mounted on the microscope. 
Cryopumping can be done with a liquid nitrogen cold finger, which encloses the sample area, allowing for local vacuum levels lower than the base pressure in the rest of the column.

Another challenge is the mitigation of electron deflection due to the radio frequency (RF) trap potentials (with amplitudes of up to $50~\si{\volt}$) needed for ion trapping. This can be achieved by synchronizing the electron transit with the trap drive. In practice, this requires that the electrons traverse the trapping region near the zero-crossing of the RF electric field, where the field amplitude is close to zero. This condition can be ensured by using, for example, ultrafast TEM setups, which allow for precise timing of the electron pulses with lengths of a few hundred femtoseconds \cite{FEIST201763}. Compared to usual trap drive frequencies of around $20~\mathrm{MHz}$, resulting in oscillation periods of $\sim\!50~\mathrm{ns}$ of the RF field, this would ensure negligible deflection of the electron wave packet.

Finally, the integration of an ion trap into a TEM requires consideration of the limited space inside the microscope. Fortunately, a microfabricated ion trap and integrated photonic components for light delivery and detection allow for a substantial reduction of the setup's space requirements. The total area that the ion trap can occupy is approximately $15~\si{\milli\meter} \times 25~\si{\milli\meter}$. Moreover, conventional pole pieces inside a TEM have a vertical spacing of just a few millimeters. Consequently, the trapped-ion platform is based on a planar surface-electrode Paul trap, in which single or multiple \ce{^{40}Ca+} ions are confined around a hundred micrometers above the chip surface [Fig.~\ref{fig:sketches} (a)]. The electrode geometry includes a sufficiently large opening around the trapping region, allowing the electron beam to pass through. This opening also enables the addressing of ions with a tailored photonic integrated chip (PIC) stacked beneath the ion trap. We envision a trap design similar to the one in reference~\cite{badawi2025chiplettechnologylargescaletrappedion}, which, with reasonable engineering adjustments, can be incorporated into a TEM via a customized TEM sample holder.  The PIC comprises grating couplers that deliver all necessary wavelengths for ion cooling, qubit manipulation, and readout. Light is supplied via fibers attached directly to the chip edge and routed through on-chip waveguides to the grating couplers \cite{Mehta2020}. While the integrated optics handle light delivery, light collection is performed separately using a fiber with a lens attached to its facet. To efficiently
detect the ion's fluorescence, we require a numerical aperture of at least $0.1$ \cite{schindler_quantum_2013}. 

\section{Scattering operator}
\label{ap:scattering_operator}
This appendix provides a detailed derivation of the scattering operator describing the interaction between free electrons and trapped ions in a TEM setup.

The scattering process depends on the proximity between the electron and the ion, with closer approaches enabling stronger coupling but potentially introducing decoherence as well. The quantum dynamics is governed by the total Hamiltonian:
\begin{equation}
    \hat{H} = \hat{H}_\text{free} + \hat{H}_\text{int}\,,
\end{equation}
where $\hat{H}_\text{free}=\hat{H}_\text{el}+\hat{H}_\text{cmi}+\hat{H}_{\rm qb}$ describes the independent evolution of both particles and $\hat{H}_\text{int}$ captures their mutual interaction.
The relativistic Hamiltonian of a free electron in atomic units $\hbar=m_{\rm e}=|e|=4\pi\epsilon_0=1$ is
\begin{equation}
    \hat{H}_\text{el} = c\sqrt{c^2 + \vectbf{\hat{p}}^2}\,,
\end{equation}
where $\vectbf{\hat{p}}$ is the momentum operator and $c$ is the speed of light. The Hamiltonian for the ion inside a harmonic trap corresponds to
\begin{equation}
    \hat{H}_\text{cmi} = \frac{\vectbf{\hat{P}}^2}{2m_\text{ion}} + \frac{1}{2}m_\text{ion}\vectbf{\hat{R}}^\text{T}\vectbf{\Omega}^T\vectbf{\Omega}\vectbf{\hat{R}}\,,
\end{equation}
where $\vectbf{\hat{P}}$ and $\vectbf{\hat{R}}$ are momentum and position operators, respectively, $m_\text{ion}$ is the ion mass, and $\vectbf{\Omega}=\diag(\Omega,\Omega,\Omega_Z)$ contains the trap frequencies. For simplicity we assume $\Omega_X=\Omega_Y=\Omega$ here.
The free Hamiltonian for the qubit encoded in an electronic quadrupole transition of the ion [the transition between S$_{1/2}$ and D$_{5/2}$ in Fig.~\ref{fig:sketches} (c)] is:
\begin{equation}
\hat{H}_{\rm qb} = \omega \hat\sigma^z_{\rm qb},
\end{equation}
with the Pauli-Z matrix $\hat \sigma^z_{\rm qb}$.
The ion and the electron are coupled in the scattering process through the Coulomb interaction
\begin{equation}
    \hat{H}_\text{int} = -\frac{1}{\abs*{\vectbf{\hat{r}} - \vectbf{\hat{R}}}}\,,
\end{equation}
which depends on the relative distance between the electron at position $\vectbf{\hat{r}}$ and the ion at position $\vectbf{\hat{R}}$.
The interaction entangles the free electron quantum state with the ion's vibrational states, creating the basis for quantum operations. 

While it is possible for a free electron to excite electronic transitions in the ion, this coupling is much weaker for all parameter regimes considered here and can thus be treated only in terms of a decoherence channel. In App.~\ref{ap:decoherence_internal} we estimate the magnitude of this effect. In the context of electron-ion scattering, we focus on the vibrational motion of the ion inside the harmonic trap, as this couples directly to the electron's electric field via the Coulomb interaction. The vibrational modes are quantized, with the zero-point motion characterized by the characteristic length scales $R_0 = \sqrt{1/(m_\text{ion}\Omega)}$ and $Z_0 = \sqrt{1/(m_\text{ion}\Omega_Z)}$. For typical trapped ions, $R_0$ and $Z_0$ are of 
the order of 1-100$~\si{\nano\meter}$, much larger than atomic scales due to the weak confinement. The ion can be prepared in various excited states, including Fock states $|n\rangle$ or, more importantly for our purposes, coherent states $|\alpha\rangle$, which closely resemble classical oscillators. These coherent states exhibit minimum uncertainty and allow us to treat the ion's position as having a well-defined classical trajectory with quantum fluctuations, making them ideal for studying the quantum aspects of the electron-ion interaction, while maintaining a semi-classical picture of the ion's motion.

To study the interaction, we employ the paraxial approximation, which is justified by several physical considerations. First, the transverse momentum distribution is centered at zero with a spread $\Delta p_\perp$ that is at least $10^{2}$ times smaller than the central longitudinal momentum $p_0$ along the $z$ axis~\cite{egerton2011_EELS_EM, spence2003high}. For momentum-independent interactions, this allows us to consider the evolution primarily along the longitudinal direction.
Second, the longitudinal momentum spread of the electron wavefunction in state-of-the-art electron microscopes is sufficiently large to render the momentum transfer to the longitudinal component negligible. For inelastic scattering processes, the momentum transfer relative to $p_0$ scales as $\Delta n \Omega \gamma(p_0) / p_0^2 \sim \Delta n 10^{-10}$ for a \SI{100}{\eV} beam (see~\cite[p.~124-125]{egerton2011_EELS_EM}), where $\gamma(p_0)$ is the relativistic gamma factor at momentum $p_0$ and $\Delta n$ is the change in vibrational quantum number of the ion, becoming even smaller for relativistic electron beams. 
For elastic scattering processes involving a very focused probe relative to the scatterer size, the relative momentum transfer for direct impact scales as $(R_0 p_0)^{-2} \sim 10^{-6}$ at \SI{100}{\eV}, decreasing further at relativistic energies. 
These momentum transfers are orders of magnitude smaller than the longitudinal momentum spread of the source. By combining the coherence length relations from~\cite[p.~113]{spence2003high} with the state-of-the-art energy spread values reported in~\cite[p.~39-42]{egerton2011_EELS_EM}, we estimate the relative longitudinal spread to range from $\Delta p_z/p_0 \sim 10^{-3}$ for \SI{100}{\eV} electrons down to $\Delta p_z/p_0 \sim 10^{-5}$ at \SI{100}{\keV}.
Consequently, the longitudinal component of the electron wavefunction remains largely unaffected during the interaction,
an effect that is even more pronounced in aloof scattering geometries. Altogether, this justifies the use of the paraxial approximation.

The scattering operator governing the electron-ion interaction is given by the standard expression 
\begin{equation}
    \hat{S} = \hat U_{\rm free}(0,\infty) \hat U(\infty,-\infty)\hat U_{\rm free}(-\infty,0),
\end{equation}
where $\hat U_{\rm free}$ and $\hat U$ are the free and full evolution operators. 
It can be evaluated in the interaction picture as
\begin{equation}
    \hat{S} = \mathcal{T} \exp \left( -i \int_{-\infty}^{+\infty} \dd t \ \hat{H}^\text{I}_\text{int}(t) \right),
\end{equation}
where $\mathcal{T}$ denotes the time-ordering operator.
The scattering process occurs in a stroboscopic regime, characterized by a separation of timescales: the electron's interaction time (typically several femtoseconds) is much shorter than the ion's dynamical timescale (nanoseconds for 3 MHz precession frequencies). This temporal hierarchy allows each electron to effectively sample a static ion configuration during its passage.

In the interaction picture, the time-dependent Hamiltonian is:
\begin{equation}
    \hat{H}^{\text{I}}_\text{int}(t) = -\frac{1}{\abs*{\vectbf{\hat{r}}^\text{I}(t) - \vectbf{\hat{R}}^{\text{I}}(t)}},
\end{equation}
with $\vectbf{\hat r}^\text{I}(t)= \vectbf{\hat r} - v_\text{el} t \vectbf{e_z}$ and $\vectbf{\hat R}^\text{I}(t)=\vectbf{\hat R}^\text{I}_\perp(t) + \hat Z^\text{I}(t)\vectbf{e_z}$,
where
$
    \vectbf{\hat R}^\text{I}_\perp(t) = \vectbf{\hat R}_\perp \cos(\Omega t)+\vectbf{\hat P}_\perp/m_\text{ion}\sin(\Omega t)
$ and
$
    {\hat Z}^\text{I}(t) = {\hat Z} \cos(\Omega_Z t)+{\hat P_Z}/m_\text{ion}\sin(\Omega_Z t)
$.
The stroboscopic approximation enables significant simplifications. First, the ion position can be treated as frozen: $\vectbf{\hat R}^{\text{I}}(t) \approx 
\vectbf{\hat R}^\text{I}(t_\text{el})
$.  
Here, $t_\text{el}$ is the time of arrival of the electron. Second, under the paraxial approximation, the electron trajectory follows $\vectbf{\hat r}^\text{I}(t) = \vectbf{\hat r} - v_\text{el} t \vectbf{e}_{\rm z}$. Crucially, in this framework the interaction Hamiltonian commutes with itself at different times, $[\hat{H}^{\text{I}}_\text{int}(t), \hat{H}^{\text{I}}_\text{int}(t')] = 0$, allowing the time-ordering operator to be dropped. This commutation arises because the time dependence enters only through the classical electron trajectory.

The scattering operator consequently takes the simplified form:
\begin{equation}
    \hat{S} = \exp \left( i\int_{-\infty}^{+\infty} \dd t \ \frac{1}{\abs*{\vectbf{\hat{r}} - v_\text{el} t \vectbf{e}_{z} - \vectbf{\hat{R}}^{\text{I}}(t_\text{el})}} \right).
\end{equation}
We now evaluate the integral in the exponent by defining the
vectorial impact parameter 
$
\vectbf{\hat{b}}(t_\text{el}) = \vectbf{\hat{r}}_\perp - \vectbf{\hat{R}}^{\text{I}}_\perp(t_\text{el})
$ with the transverse free electron position $ \vectbf{\hat{r}}_\perp=(\hat{x},\hat{y},0)$. With the longitudinal position $ \hat{z}$ and $\hat Z^{\rm (I)}(t_{\rm el})$ for the free electron and ion, respectively,
the time integral becomes:
\begin{align}
\hat{\tau} &= \int_{-\infty}^{+\infty} \frac{\dd t}{\abs*{\vectbf{\hat{b}}(t_\text{el}) + (\hat z - \hat Z^\text{I}(t_\text{el}) - v_\text{el} t )\vectbf{e_z}}} \\\nonumber
&= \frac{1}{v_{\rm el}}\int_{-\infty}^{+\infty} \frac{\dd u}{\sqrt{\hat{b}(t_\text{el})^2+u^2}} \\\nonumber
&= \frac{2}{v_{\rm el}}\lim_{L\rightarrow\infty}
\text{arsinh}\left(\frac{L}{\hat{b}(t_\text{el})}\right),
\end{align}
where we have introduced the substitution $u = \hat{z}-\hat{Z}^{\text{I}}(t_\text{el})-v_{\rm el}t$ and defined the magnitude of the impact parameter operator $\hat{b}(t_\text{el}) = |\vectbf{\hat{b}}(t_\text{el})|$.
Upon using the asymptotic expansion $\text{arsinh}(z) = \log(2z) + \mathcal{O}(z^{-2})$, we get:
\begin{equation}
    \hat \tau = \frac{2}{v_\text{el}} \lim_{L\to\infty}\log(\frac{2L}{\hat b(t_\text{el})}).
\end{equation}

The scattering operator thus comprises two distinct contributions: a divergent global phase and a finite operator-valued phase, which depends
on the impact parameter $\hat{b}$.
The apparent divergence arises from the idealized treatment of an infinite interaction range, but physical implementations naturally regularize this through charge screening. The electron beam possesses a finite transverse extent (typically microns), and the interaction length is bounded by the ion trap dimensions (hundreds of microns). This physical confinement ensures that electrons with impact parameters exceeding the interaction length do not contribute measurably. Consequently, the divergent term becomes a state-independent global constant, while the $\hat{b}$-dependent term captures the essential physics of the interaction.

When computing observable quantities, such as interference patterns or state transition probabilities, only the phase difference at different impact parameters is relevant. In these relative measurements, thus, the divergent global phase cancels exactly, leaving well-defined finite results.
The physically relevant scattering operator is therefore:
\begin{equation}\label{eq:S_final}
    \hat{S} = \exp \left( -\frac{2i}{v_\text{el}} \log(\hat{b}(t_\text{el})) \right),
\end{equation}
which describes the coherent phase imprinted onto the combined electron-ion wavefunction induced by the Coulomb interaction. Upon inserting $\hat b(t_\text{el}) = \abs*{\vectbf{\hat r}_\perp - \vectbf{\hat R}_\perp^\text{I}(t_\text{el})}$, this becomes Eq.~\eqref{eq:scattering_operator}.

\section{Explicit computations on the action of the scattering operator on coherent states}
\label{ap:explicit_computation_of_phase}
In this appendix, we explicitly evaluate the displacement dependence of the phase shift that the joint electron-ion wavefunction acquires through the Coulomb interaction and also compute the probability for the state to change by more than a phase.
For this we start from Eq.~\eqref{eq:scattering_operator}
and evaluate the matrix element
\begin{equation}
\mathcal{S}(\vectbf{b},\boldsymbol{\upalpha})=
\bra{\psi_\vectbf{b}}_\text{el}\bra{\boldsymbol{\upalpha}}_\text{cmi}\hat S\ket{\boldsymbol{\upalpha}}_\text{cmi}\ket{\psi_\vectbf{b}}_\text{el}\,,    
\end{equation}
with the incoming scattering states $\ket{\boldsymbol{\upalpha}}_\text{cmi}$ and $\ket{\psi_\vectbf{b}}_\text{el}$. The state $\ket{\psi_\vectbf{b}}_\text{el}$ is a Gaussian state, that, when propagated freely from $t=-\infty$ to $t=0$, ends up focused in the plane of the ion at position $\vectbf{b}$ with a spread of $\delta_{\vectbf{r_\perp}}$:
\begin{equation}
    \label{eq:gaussian_probe}
    \braket{\vectbf{r}_\perp}{\psi_\vectbf{b}} = (2\pi \delta_{\vectbf{r_\perp}}^2)^{-1/2}e^{-\frac{(\vectbf{r}_\perp-\vectbf{b})^2}{4\delta_{\vectbf{r_\perp}}^2}}.
\end{equation}
The phase shift $\Delta\phi$ is then given by
\begin{equation}
    \Delta\phi = \arg(\mathcal{S}(\vectbf{b},\boldsymbol{\upalpha})),
\end{equation}
where $\arg$ denotes the complex argument function. The probability for the state to change by more than a phase is
\begin{equation}
    P_{\rm scatt} = 1-\abs*{\mathcal{S}(\vectbf{b},\boldsymbol{\upalpha})}^2.
\end{equation}
Since 
\begin{equation}
    \abs*{\braket*{\vectbf{R}_\perp^\text{I}(t_\text{el})}{\boldsymbol{\upalpha}}}^2
    =
    \frac{1}{\pi R_0^2}e^{-\left(\frac{\vectbf{R}_\perp}{R_0}-\sqrt{2}\Re(\boldsymbol{\upalpha}e^{i\Omega t_\text{el}})\right)^2},
\end{equation}
we can write the scattering matrix element $\mathcal{S}(\vectbf{r}_\perp,\boldsymbol{\upalpha})$ 
in terms of the two-dimensional convolution of the function 
$S(\vectbf{x}) = \abs*{\vectbf{x}}^{-2i/v_\text{el}}$
with two radially symmetric Gaussians
$G_{a}(\vectbf{x}) = (\pi a^2)^{-1}e^{-{\vectbf{x}^2}/{a^2}}$ parametrized by their radial widths $a$:
\begin{equation}
    \mathcal{S}(\vectbf{b},\boldsymbol{\upalpha})=
    (G_{\sqrt{2}\delta_{\vectbf{r_\perp}}} \kern-1em \ast G_{R_0} \!\! \ast \!S)(
    \vectbf{b}
    \!-\!
    \sqrt{2} R_0 \Re(\boldsymbol{\upalpha}e^{i\Omega t_\text{el}})
    )\,.
\end{equation}
We can combine the two Gaussian functions making use of the relation
$
    G_{\sqrt{2}\delta_{\vectbf{r_\perp}}} \!\!\!\!\ast G_{R_0} = G_{R_{\rm eff}}
$, with $R_{\rm eff} = (R_0^2+2\delta_{\vectbf{r_\perp}}^2)^{1/2}$.
It proves useful to define the convolution as a new function $\Sigma_a=G_a\ast S$.
Then
\begin{equation}
    \mathcal{S}(\vectbf{b},\boldsymbol{\upalpha})=
    \Sigma_{R_{\rm eff}}(\vectbf{b}
    -
    \sqrt{2} R_0 \Re(\boldsymbol{\upalpha}e^{i\Omega t_\text{el}})
    ).
\end{equation}
We can show that the function $\Sigma_a$ has a representation in terms of the Kummer confluent hypergeometric function $_1F_1$ and Euler's Gamma function:
\begin{equation}
\label{eq:analytic_form_of_sigma}
    \Sigma_a(\vectbf{b}) = 
    \frac{
        \Gamma\left(1-\frac{i}{v_\text{el}}\right)
    }{
        a^{\frac{2i}{v_\text{el}}}
    }
    \;{_1F_1}\left(\frac{i}{v_\text{el}},1,-\frac{\vectbf{b}^2}{a^2}\right).
\end{equation}
In order to prove this, write the convolution with the help of the Fourier transform:
\begin{equation}
    \mathcal{F}[\Sigma_a]=\mathcal{F}[G_a]\mathcal{F}[S].
\end{equation}
The Fourier transform of the Gaussian is again a Gaussian: 
$
\mathcal{F}[G_a](\vectbf{k}) = 
e^{-\frac{a^2}{4}\vectbf{k}^2}
$.
The Fourier Transform of $S$ can be evaluated to:
\begin{equation}
    \mathcal{F}[S](\vectbf{k})=
    2^{1-\frac{2i}{v_{\text{el}}}}
    \frac{
    \Gamma\left(
    1-\frac{i}{v_\text{el}}
    \right)}
    {
    \Gamma\left(
    \frac{i}{v_\text{el}}
    \right)
    }
    \abs*{\vectbf{k}}^{-2(1-\frac{i}{v_\text{el}})}
.
\end{equation}
Since both functions are radially symmetric, we can directly evaluate the angular integral in the inverse Fourier transform to obtain:
\begin{equation}
    \Sigma_a(\vectbf{b})=c
    \int_0^\infty k^{\frac{2i}{v_\text{el}}-1}e^{-\frac{a^2}{4}k^2} J_0(k\abs*{\vectbf{b}}) \dd{k},
\end{equation}
with 
$c = 2^{1-\frac{2i}{v_{\text{el}}}}
    \Gamma(
    1-\frac{i}{v_\text{el}}
    )
    \Gamma(
    \frac{i}{v_\text{el}}
    )^{-1}
    $.
This integral is listed in \cite{gradshteyn2007} (entry 6.631.1) and gives:
\begin{equation}
    \Sigma_a(\vectbf{b})=c\frac{\Gamma\left(\frac{i}{v_\text{el}}\right)}{2\left(\frac{a^2}{4}\right)^{\frac{i}{v_\text{el}}}}
    {_1F_1}\left(\frac{i}{v_\text{el}},1,-\frac{\vectbf{b}^2}{a^2}\right)\,.
\end{equation}
Upon canceling the coefficients we end up with Eq.~\eqref{eq:analytic_form_of_sigma}.

For $\vectbf{b}=0$, we can explicitly simplify the absolute square of $\Sigma_a(\vectbf{b})$. Using ${_1F_1}(\nu,\mu,0)=1$, $\overline{\Gamma(z)}=\Gamma(\bar z)$ together with $\Gamma(1+z)=z\Gamma(z)$ and $\Gamma(1-z)\Gamma(z)=\pi/\sin(\pi z)$, we find:
\begin{align}
    \abs*{\Sigma_a(\vectbf{0})}^2
    &= \abs{\Gamma\left(1-\frac{i}{v_\text{el}}\right)}^2 \\\nonumber
    &= 
    \frac{i}{v_\text{el}}\Gamma\left(1-\frac{i}{v_\text{el}}\right)\Gamma\left(\frac{i}{v_\text{el}}\right) \\\nonumber
 &= 
    \frac{i}{v_\text{el}}\frac{\pi}{\sin(\frac{i\pi}{v_\text{el}})} = 
    \frac{\pi}{v_\text{el}\sinh(\frac{\pi}{v_\text{el}})}.
\end{align}

\section{Back action on a focused probe}
\label{ap:backaction}
In the main text, we argued that the back action on an electron in a very focused probe may be ignored. In this appendix, we give a derivation of this fact and show where this assumption breaks down.

We study the scattering probability of the electronic state, assuming the ionic state to remain unchanged. We again assume the initial state of the electron to be a Gaussian [Eq.~\eqref{eq:gaussian_probe}] in the transverse plane and further assume the ion to be in the ground state. The analogous argument applies to coherent states since the interaction depends only on the impact parameter. The scattering probability as a function of $\abs*{\vectbf{b}}$ can be evaluated as
\begin{equation}
    \label{eq:electronic_overlap}
    \eta(\abs*{\vectbf{b}}) = 1-\frac{\abs*{\bra{\psi_\vectbf{b}}_\text{el}\bra{0}_\text{cmi}{\hat S}\ket{0}_\text{cmi}\ket{\psi_\vectbf{b}}_\text{el}}^2}{\int \dd^2{r_\perp} \; \abs*{\bra{\vectbf{r_\perp}}_\text{el}\bra{0}_\text{cmi}{\hat S}\ket{0}_\text{cmi}\ket{\psi_\vectbf{b}}_\text{el}}^2}\,,
\end{equation}
We can write the overlap $\eta$ in terms of a fraction of convolutions:
\begin{equation}
    \eta = 1-
    \frac{
        \abs*{(G_{\sqrt{2}\delta_{\vectbf{r_\perp}}}\ast G_{R_0} \ast S)}^2
    }{
        (\abs*{G_{R_0}\ast S}^2 \ast G_{\sqrt{2}\delta_{\vectbf{r_\perp}}})
    }\,,
\end{equation}
where $G_{a}(\vectbf{x}) = (\pi a^2)^{-1}e^{-\frac{\vectbf{x}^2}{a^2}}$, and $S(\vectbf{x}) = \abs*{\vectbf{x}}^{-\frac{2i}{v_\text{el}}}$. With
$
    \Sigma_{a} = G_{a} \ast S
$ (compare App.~\ref{ap:explicit_computation_of_phase}),
\begin{figure}
    \includegraphics[width=\linewidth]{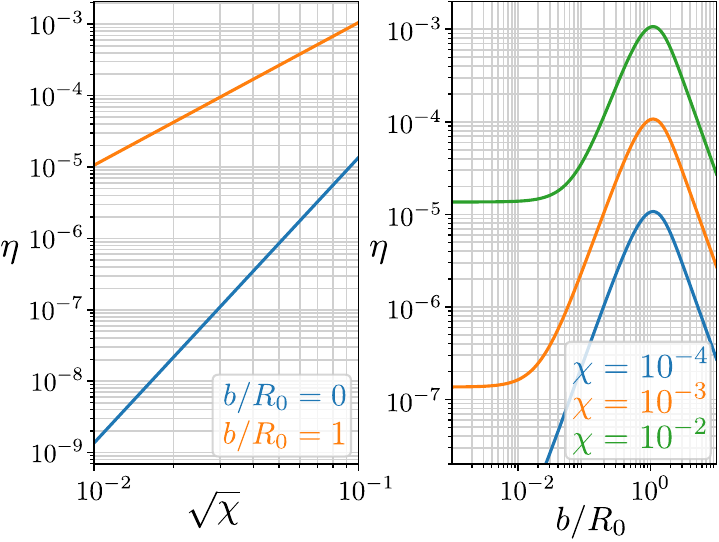}
    \caption{The change in the final electronic state for a focused electron beam of 100 eV ($v_\text{el}\sim 3$) measured in terms of $\eta$ (Eq.~\eqref{eq:electronic_overlap}) as a function of the relative probe size $\sqrt{\chi}=\sqrt{2}\delta_{\vectbf{r_\perp}}/R_0$ and the impact parameter $b =\abs*{\vectbf{b}}$.}
    \label{fig:decoherence_due_to_electron_size}
\end{figure}
we get:
\begin{align}
    \label{eq:eta_formula}
    \eta &= 1-\frac{
        \abs*{\Sigma_{R_0} \ast G_{\sqrt{2}\delta_{\vectbf{r_\perp}}}}^2
    }{
        \abs*{\Sigma_{R_0}}^2 \ast G_{\sqrt{2}\delta_{\vectbf{r_\perp}}}
    }
    \nonumber\\ &=
    \frac{
        \abs*{\Sigma_{R_0}}^2 \ast G_{\sqrt{2}\delta_{\vectbf{r_\perp}}}
        -
        \abs*{\Sigma_{R_0} \ast G_{\sqrt{2}\delta_{\vectbf{r_\perp}}}}^2
    }{
        \abs*{\Sigma_{R_0}}^2 \ast G_{\sqrt{2}\delta_{\vectbf{r_\perp}}}
    }.
\end{align} 
In Eq.~\eqref{eq:analytic_form_of_sigma} we give an explicit formula for the function $\Sigma_{R_0}$.
As the prefactors in Eq.~\eqref{eq:analytic_form_of_sigma} appear in the numerator and the denominator of Eq.~\eqref{eq:eta_formula} and therefore cancel out, it is useful to define
$
    u(s) = \;_1F_1(
        {i}/{v_\text{el}},
        1,
        -s
    )
$, such that then
$
    \Sigma_{R_0}(\vectbf{x}) = \Gamma(1-i/v_\text{el}) R_0^{-2i/v_\text{el}}u(\vectbf{x}^2 / R_0^2)
$.
For a narrow beam with
$
\chi:=2\delta_{\vectbf{r_\perp}}^2/R_0^2 \ll 1$,
we may work with the operator representation 
$
    G_{\sqrt{2}\delta_{\vectbf{r_\perp}}} \ast \cdot = 
    \exp(\chi R_0^2/4 \; \Delta)
$, where $\Delta$ denotes the Laplacian operator in two dimensions. This representation follows from a Fourier transform and is exact as long as the exponential series converges, which, in our case, is ensured for $\chi < 1$. In order to see this, note that 
$
  (2\pi)^2\norm{(R_0^2/4)^n\Delta^n\Sigma_{R_0}}_\infty \leq 
  \norm*{(R_0\abs*{\vectbf{k}}/2)^{2n}\mathcal{F}[\Sigma_{R_0}]}_1
  \sim \Gamma(n)
$. The factorial growth of the derivatives cancels out with the factorial in the denominator of the exponential series, resulting in the convergence radius of $1$. In terms of the variable $s=\vectbf{x}^2/R_0^2$, and acting on radially symmetric functions, the convolution with the Gaussian becomes:
$
G_{\sqrt{2}\delta_{\vectbf{r_\perp}}} \ast \cdot = \exp(\chi( s\partial_{s}^2+\partial_{s}))
$.
With this, we get:
\begin{align}
    \label{eq:eta_formula_in_u}
    \eta &=
    \frac{
        e^{\chi ( s\partial_{s}^2+\partial_{s})} \abs*{u}^2
        -
        \abs*{ e^{\chi ( s\partial_{s}^2+\partial_{s})} u}^2
    }{
        e^{\chi ( s\partial_{s}^2+\partial_{s})} \abs*{u}^2
    }.
\end{align}
In the following, we expand Eq.~\eqref{eq:eta_formula_in_u} to leading order in $\chi\to 0$. For $s=0$, the numerator approaches zero quadratically as $\chi\to 0$. The denominator stays finite and greater than zero. Thus, the leading order, in this case, amounts to an expansion of the numerator to second order and of the denominator to zero'th order. For $s>0$, the root of the numerator is linear and a linear expansion of the numerator would suffice. In order to capture also the behavior for small $s\to0^+$, however, we expand the numerator to second order also in this case, as this does not change the validity of the leading order expansion. We drop, however, any second order terms that become zero for $s\to 0$. We get
\begin{align}
\label{eq:eta_approximation}
    \abs*{u}^2\eta \approx
    2s\chi \abs*{u'}^2
    + \chi^2\abs*{u'}^2.
\end{align}
Using the general relation
$
    \partial_z \; {_1F_1}(\alpha,\beta,z) = \frac{\alpha}{\beta} {_1F_1}(\alpha+1,\beta+1,z)
$,
we can evaluate the derivative $u'(s)= -i/v_{\rm el} \, _1F_1(1+i/v_{\rm el},2,-s)$.
The expression Eq.~\eqref{eq:eta_approximation} for the scattering probability $\eta$ is visualized in Fig.~\ref{fig:decoherence_due_to_electron_size}. As it is apparent, it practically vanishes for small impact parameters. It peaks at $b=R_0$, where the gradient of the ionic charge density is the largest. The maximal scattering probability is of the order of $\sim\!10^{-3}$ for an electron energy of $100~\si{\eV}$ and is, to a good approximation, inversely proportional to the electron energy. Thus, for electron energies of $1~\si{\kilo\eV}$, $10~\si{\kilo\eV}$ and $100~\si{\kilo\eV}$ we find a maximum of $10^{-4}$, $10^{-5}$ and $10^{-6}$ accordingly. For $b=0$, we find $\eta \approx \chi^2\abs*{u'(0)}^2 / \abs*{u(0)}^2 = \chi^2 / v_\text{el}^2$.
Note that the scaling is quadratic in $v_\text{el}$ rather than quartic as one might expect from Rutherford scattering. The difference from Rutherford scattering comes from the fact that we consider a strongly focused probe with a well-defined impact parameter instead of plane waves.

\section{Probability of internal excitations of the trapped ion}
\label{ap:decoherence_internal}
In the main text, as well as in App.~\ref{ap:scattering_operator}, only the coupling of the electron to the center-of-mass motion of the trapped ion was included. The degrees of freedom in the electronic subsystem of the trapped ion were thereby neglected. 
Using theoretical and experimental data for the total scattering cross section of electrons with atomic potassium and atomic calcium, we estimate the decoherence that arises from electron-induced excitations in the electronic subsystem of the trapped ion.
The total (elastic and inelastic) scattering cross sections reported~\cite{Kwan1991, Msezane1992, Ratnavelu2011} for potassium and an impact energy of $100~\si{\eV}$ are in the range of $\sigma_{\rm tot} ~/~ (a_0^2\pi) \sim 28-36$, where $a_0$ is the Bohr radius, while for calcium~\cite{Raj2007,Kelemen1995} they are in the range of $\sigma_{\rm tot} ~/~ (a_0^2\pi) \sim 15-25$. From this, assuming the ion to be in a coherent state and neglecting the spread of the incident electron beam, we estimate the scattering probability to be
\begin{equation}
    P_{\rm scatt} \lesssim \int_0^{\sqrt{\sigma_{\rm tot}/\pi}}
    \frac{R_\perp\dd{R_\perp}}{\pi R_0^2}e^{-\frac{R_\perp^2}{R_0^2}} \approx \frac{\sigma_{\rm tot}}{\pi R_0^2},
\end{equation}
which, with $\sigma_{\rm tot} \lesssim 36a_0^2\pi$, results in $P_{\rm scatt} \lesssim 2\cdot 10^{-5}-2\cdot 10^{-4}$ for $R_0=40~\si{\nano\meter}$ and $R_0=13~\si{\nano\meter}$ respectively. These probabilities are well below the decoherence due to state changes in the ionic center of mass (see the main text as well as App.~\ref{ap:explicit_computation_of_phase}) and comparable to the decoherence due to the finite extent of the electronic beam profile and the back action onto the free electron arising from it (see App.~\ref{ap:backaction}). The scattering probability decreases further for higher incident energies.

\section{Transferring the phase mask of a specimen to the quantum computer}\label{sec: phase mask}
\label{ap:phase_kick_back}
In this appendix, we explain how phase information about a weak-phase object can be transferred to the quantum computer. A related procedure can be found in \cite{okamoto_universal_2023}, though the interaction differs. We start from the one-electron-many-ion coupling Eq.~\eqref{eq:many_qubit_scattering_operator}, which we reproduce here for convenience:
\begin{equation}
    \label{eq:ap_many_qubit_scattering_operator}
    \hat U_{e-N_{\rm qb}}= 
    \sum_{k=1}^N
    e^{\frac{i}{2}g\hat\sigma^x_{{\rm qb}, k}}\ketbra*{\psi_k}_\text{el}.
\end{equation}
Note that we dropped the global phase $\kappa$ for simplicity.
Here we assumed the electron to be restricted to $N$ discrete paths denoted as $\ket*{\psi_{k}}_\text{el}$, each focused to a spot $\vectbf{r}_{\perp,k}$ in the proximity of a single corresponding ion. This may be achieved by using a diffraction grating and imaging the discrete diffraction image onto the plane where the quantum computer is inserted. 
We assume the initial state of the electron to be a coherent weighted superposition of all possible paths. We further assume the quantum computer in the form of $N$ ions to be initialized in the state $|0\ldots0\rangle_{\rm qb}$. The initial joint state reads:
\begin{equation}
    \ket{\Psi_0} = \sum_{k=1}^N w_k\ket*{\psi_{k}}_\text{el}\ket{0\ldots 0}_{\rm qb}.
\end{equation}
Note that we dropped the motional degree of freedom of the ionic center of mass here, as the final unitary Eq.~\eqref{eq:ap_many_qubit_scattering_operator} does not include it anymore.
Applying Eq.~\eqref{eq:ap_many_qubit_scattering_operator} to this joint initial state gives:
\begin{equation}
    \ket{\Psi_1} = \sum_{k=1}^N w_ke^{\frac{i}{2}g\sigma^x_{{\rm qb}, k}}\ket*{\psi_{k}}_\text{el}\ket{0\ldots 0}_{\rm qb}.
\end{equation}
The $N$ electron paths are further imaged onto the sample plane, where they interact with the specimen and acquire a path-dependent phase $\varphi_k$. After this, the joint state becomes:
\begin{equation}
    \ket{\Psi_2} = \sum_{k=1}^N w_ke^{\frac{i}{2}g\sigma^x_{{\rm qb}, k}}e^{i\varphi_k}\ket*{\psi_{k}}_\text{el}\ket{0\ldots 0}_{\rm qb}
\end{equation}
Finally, the electron is detected in momentum space. If the measurement result is $\vectbf{p}_\perp$, the wavefunction becomes:
\begin{equation}\label{eq:post_e_meas}
    \ket{\Psi_3} = \sum_{k=1}^N w_ke^{\frac{i}{2}g\sigma^x_{\rm qb}}e^{i\varphi_k}\underbrace{\braket*{\vectbf{p}_\perp}{\psi_{k}}_\text{el}}_{\propto e^{-i\vectbf{p}_\perp\cdot\vectbf{r}_{\perp,k}}}\ket{0\ldots 0}_{\rm qb}.
\end{equation}
In order to restore coherence, one has to correct for the additional phase factor $e^{-i\vectbf{p}_\perp\cdot\vectbf{r}_{\perp,k}}$. This can be done by applying the unitary
\begin{equation}\label{eq: phase_correction}
    \hat U_{\rm corr}=\prod_{k=1}^N e^{i\vectbf{p}_\perp\cdot\vectbf{r}_{\perp,k} \ketbra{1}{1}_{\rm qb, k}}
\end{equation}
to the quantum computer. With this, the final state in the quantum computer becomes:
\begin{align}
    \ket{\Psi_{\rm fin}} =& 
    \cos(g/2)\sum_{k=1}^N 
    w_k e^{i\varphi_k}
    e^{-i\vectbf{p}_\perp\cdot\vectbf{r}_{\perp,k}}
    \ket{0\ldots0}_{\rm qb}
    \nonumber \\
    &+
    \sin(g/2)\sum_{k=1}^N
    w_k e^{i\varphi_k}
    \ket*{\underbrace{0\ldots 1\ldots 0}_{k^\text{th}\text{ bit is 1}}}_{\rm qb}.
\end{align}
The phases $\varphi_k$ are then encoded as relative phases between the states $\ket{10\ldots 0}$, $\ket{01\ldots 0}, \ldots, \ket{0\ldots 01}$. From here, quantum optimal measurements can be performed in order to extract as much information as possible about whatever observable related to the phase shifts.

\section{Quantum advantage in coherent phase estimation}\label{ap:example1}
In this appendix, we demonstrate that the coherent coupling of the probe electrons to the quantum computer allows us to surpass the standard quantum limit in a standard phase estimation problem.
We first discuss an idealized setting as proposed in \cite{okamoto_entanglement-assisted_2014}, in which it is possible to reach Heisenberg scaling and explain how it can be realized in the present setup. We then generalize this to more realistic assumptions and show that, even under these conditions, the setup may provide us a quantum advantage.

We assume a two-path setup with a trapped ion situated in one of the two paths. Initially, the trapped ion is prepared in $\frac{1}{\sqrt{2}}(e^{i\beta_0}\ket{+}_{\rm qb}+\ket{-}_{\rm qb})$. Each electron is prepared in a superposition of the two paths $\frac{1}{\sqrt{2}}(\ket{\psi_{i}}_\text{el}+\ket{\psi_{s}}_{\rm el})$, which are focused to corresponding spots $\vectbf{r}_{\perp,i/s}$ far enough apart such that $\braket{\psi_{s}}{\psi_{i}}_{\rm el}\approx0$.  
In the first path, $\ket{\psi_{i}}_{\rm el}$, the electron interacts with the trapped ion via the unitary $\hat U=e^{i\kappa}e^{\frac{i}{2}g\hat \sigma^x_{\rm qb}}$ (see Eq.~(\ref{eq:qubit_scattering_operator})). In the idealized setting, we assume $g=\pi$, which could be achieved in a low-energy electron microscopy (LEEM) setup, with electron energies of the order of $100~\si{\eV}$ (see Fig.~\ref{fig:interaction_plot}). To gain intuition, recall that this unitary acts on the ion qubit states as $e^{i\kappa}e^{\frac{i}{2}\pi\hat \sigma^x_{\rm qb}}\ket{\pm}_\text{qb}=\pm ie^{i\kappa}\ket{\pm}_\text{qb}$ thus only imprinting a phase. We can now associate this phase with the accompanying electron state and view the interaction as a phase acquired by the electron rather than by the ion. Consequently, the state after initialization and electron-ion interaction becomes:
\begin{align}
    \ket{\Psi_1}=&\frac{1}{2}\big[(e^{i\beta_0}\ket{+}_{\rm qb}\!
    +\!\ket{-}_{\rm qb})\ket{\psi_{s}}_{\rm el}\!\nonumber\\
    &+\!(e^{i\beta_0}\ket{+}_{\rm qb}\!
    -\!\ket{-}_{\rm qb})ie^{i\kappa}\ket{\psi_{i}}_{\rm el}\!\big]\nonumber\\
    =&\frac{1}{2}\big[e^{i\beta_0}\ket{+}_{\rm qb}\left(\ket{\psi_{s}}_{\rm el}+ie^{i\kappa}\ket{\psi_{i}}_{\rm el}\right)\nonumber\\
    &+\ket{-}_{\rm qb}\left(\ket{\psi_{s}}_{\rm el}-ie^{i\kappa}\ket{\psi_{i}}_{\rm el}\right)\big].
\end{align}
Next, both electron paths pass through a beam splitter as demonstrated in \cite{yasin_path-separated_2018, turner_interaction-free_2021}, such that ${\frac{1}{\sqrt{2}}(\ket{\psi_{s}}_{\rm el}+ie^{i\kappa}\ket{\psi_{i}}_{\rm el})\mapsto\ket{\psi_{s}}_{\rm el}}$ and ${\frac{1}{\sqrt{2}}(\ket{\psi_{s}}_{\rm el}-ie^{i\kappa}\ket{\psi_{i}}_{\rm el})\mapsto \ket{\psi_{i}}_{\rm el}}$, resulting in the wavefunction
\begin{equation}
\ket{\Psi_2}=\tfrac{1}{\sqrt{2}}\big(e^{i\beta_0}\ket{+}_{\rm qb}\ket{\psi_{s}}_{\rm el}+\ket{-}_{\rm qb}\ket{\psi_{i}}_{\rm el}\big).    
\end{equation}
At the $\psi_{s}$ end of the beam splitter, we place a specimen, which is assumed to be a weak phase object that imprints $e^{i\phi}$ onto the electronic wavefunction. We get:
\begin{align}
    \ket{\Psi_3} = \tfrac{1}{\sqrt{2}}\big(e^{i(\beta_0+\phi)}\ket{+}_{\rm qb}\ket{\psi_{s}}_{\rm el}+\ket{-}_{\rm qb}\ket{\psi_{i}}_{\rm el}\big)
\end{align}
Then, the electron is measured in momentum space. The overlap between any momentum state $\ket{\vectbf{p}_\perp}_{\rm el}$ with any of the two beams is $\braket{\vectbf{p}_\perp}{\psi_{i/s}}_{\rm el} \propto e^{-i\,\vectbf{p}_\perp\cdot\vectbf{r}_{\perp,i/s}}$. Taking into account that we only have two paths, we define the phase difference $\xi(\vectbf{p}_\perp)=\vectbf{p}_\perp\cdot(\vectbf{r}_{\perp,s}-\vectbf{r}_{\perp,i})$.
\begin{equation}
    \ket{\Psi_4}=\tfrac{1}{\sqrt{2}}\big(e^{i(\beta_0+\phi-\xi(\vectbf{p}_\perp))}\ket{+}_{\rm qb}+\ket{-}_{\rm qb}\big)\label{eq:example_step_4}
\end{equation}
Depending on the measurement outcome $\vectbf{p}_\perp$, we apply the correction 
$
    \hat U=e^{\frac{i}{2}\xi(\vectbf{p}_\perp)\hat \sigma^x_{\rm qb}}
$ 
to the ion, leading to
\begin{equation}
    \ket{\Psi_5}=\tfrac{1}{\sqrt{2}}\big(e^{i(\beta_0+\phi)}\ket{+}_{\rm qb}+\ket{-}_{\rm qb}\big).\label{eq:example_step_5}
\end{equation}
After the first electron, we obtain $\beta_1=\beta_0+\phi$, and likewise after the $j$-th electron, we have $\beta_{j+1}=\beta_j+\phi$. After a total of $n$ electrons, this accumulates to $\beta_n=\beta_0+n\phi$. We choose to initialize in $\beta_0=0$, and thus the final ion state will be
\begin{align}
    \ket{\Psi_{\rm fin}}_{\rm qb} = \tfrac{1}{\sqrt{2}}\big(e^{in\phi}\ket{+}_{\rm qb}+\ket{-}_{\rm qb}\big), \label{eq:final_qb}
\end{align}
which paves the way to a metrological advantage in the estimation of $\phi$. One can think of this procedure as essentially performing a multi-pass scheme without sending the electron back and forth. That is, in both cases, the idea is to coherently add up the phases imprinted by the specimen.

To round up, we quantify the metrological advantage by looking at the canonical metric of the Fisher information. The probability that a measurement in the computational basis of the final state given in Eq.~\eqref{eq:final_qb} has the result $|0\rangle_{\rm qb}$ is $p_0(\phi)=\left|\langle0|\Psi_{\rm fin}\rangle\right|^2=\frac{1}{4}\left|e^{in\phi}+1\right|^2 = \cos^2\frac{n\phi}{2}$. The probability to measure the outcome $|1\rangle_{\rm qb}$ is therefore $p_1(\phi)=1-p_0(\phi)=\sin^2\frac{n\phi}{2}$. 
With these two possible measurement probabilities, we can calculate the Fisher information as
\begin{align}
    F(\phi)&= p_0(\phi)\,(\partial_{\phi} \log p_0(\phi))^{\!2} + p_1(\phi)\,(\partial_{\phi} \log p_1(\phi))^{\!2} \nonumber\\
    &= \frac{\left(\partial_{\phi} p_0(\phi)\right)^2}{p_0(\phi)\,(1-p_0(\phi))}=n^2.\label{eq: formula_FI}
\end{align}
The task of phase estimation is widely explored in the literature, and for $n$ queries of the specimen, a Fisher information of $n^2$ is known to be the best possible outcome~\cite{Giovannetti_2006}. 

In reality, not every electron will be successfully detected. This causes significant problems in the presented scheme, because the ion and electron form a maximally entangled state. Thus, losing a single electron creates a maximally mixed state on the ion, erasing all acquired phase information and requiring a restart of the procedure. Here, we briefly analyze that, under modest loss rates, we can still gain an advantage on average.

\begin{figure}
    \includegraphics[width=\linewidth]{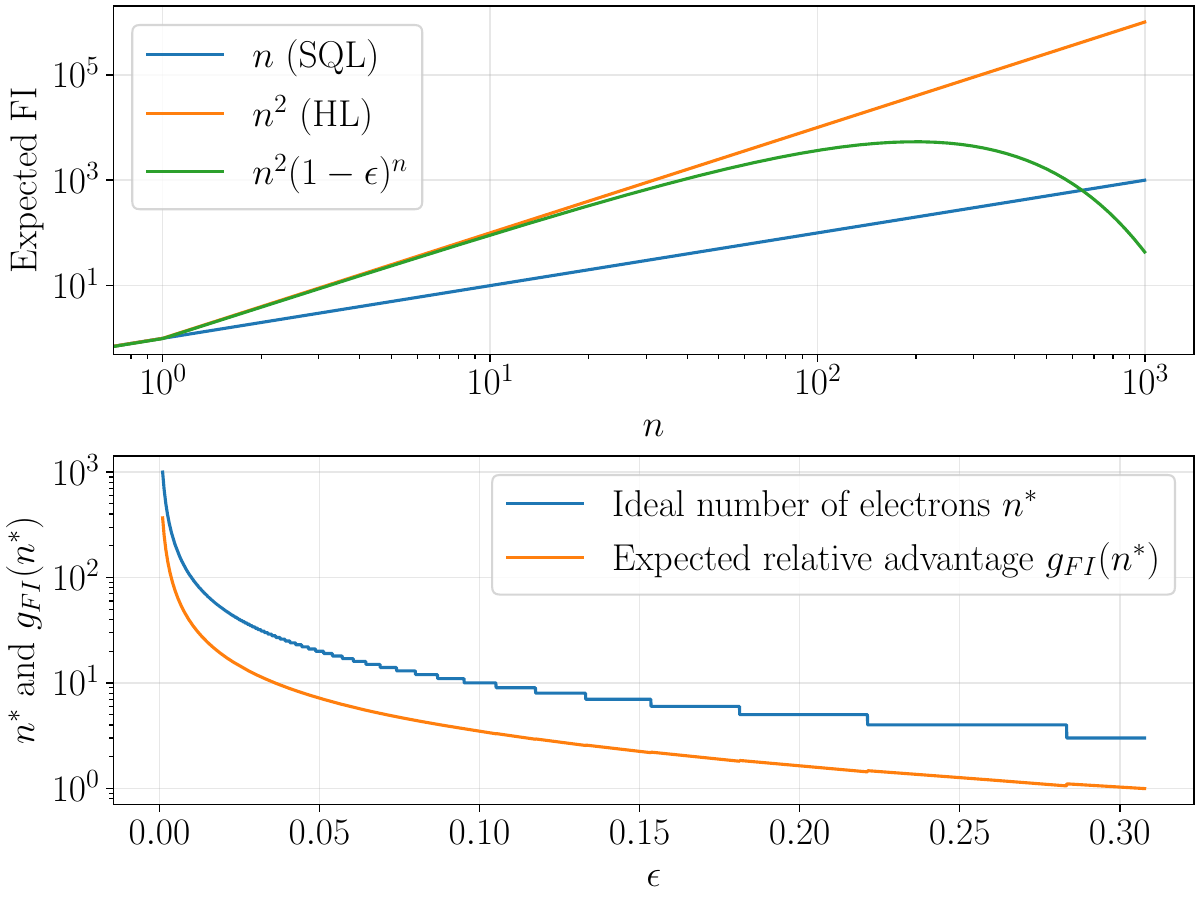}
    \caption{Top: Comparison of the expected $F(\phi)$ given by $n^2(1-\epsilon)^n$ for $\epsilon=0.01$ to the SQL $n$ and the HL $n^2$ as a function of the number of electrons $n$.
    Bottom: Depending on the probability of electron loss $\epsilon$, we show two functions. First, the ideal number of electrons, which is given by rounding $n^*=\frac{-1}{\log(1-\epsilon)}$. Second, inserting this value back into the expected ratio of obtained $F(\phi)$ and SQL, $g_{\text{FI}}(n^*)$, which quantifies the quantum advantage.
    }
    \label{fig:electron loss}
\end{figure}

Assume a probability $\epsilon$ that an electron is lost. Then, the probability of detecting all $n$ electrons is $(1-\epsilon)^n$. In this case, the Fisher information is $F(\phi)=n^2$. In all other cases, maximal mixedness has occurred at some point, and we get $F(\phi)=0$. The expected Fisher information becomes $\mathbbm{E}(F(\phi))=n^2(1-\epsilon)^n$. Fig.~\ref{fig:electron loss} (top) compares this function to the standard quantum limit (SQL) $F(\phi)=n$ and the Heisenberg limit (HL) $F(\phi)=n^2$.

We want to optimize the relative metrological advantage given by $g_{\text{FI}}(n)\coloneqq\frac{\mathbbm{E}(F(\phi))}{\text{SQL}}=n(1-\epsilon)^n$. This function has a unique maximum at $n^*=\frac{-1}{\log(1-\epsilon)}$ with a value of $g_{\text{FI}}(n^*)=\frac{-1}{e\log(1-\epsilon)}$. This ideal $n^*$, rounded to the nearest integer, as well as the corresponding expected relative gain, are shown in Fig.~\ref{fig:electron loss} (bottom) as a function of $\epsilon$. On average, an advantage is possible up to $\epsilon\approx0.3$.

Finally, we will now briefly discuss the example of coherent phase estimation with a non-ideal coupling strength $g$. In principle, all calculations can be easily repeated up to the compensation of the measurement result, Eq.~\eqref{eq:example_step_5}. This compensation applies only to $g=\pi$ (maximal entanglement), for which it was tailored. 

The other extreme is $g=0$ (no entanglement), in which case the ion state is completely unchanged. Thus, there is also no need to apply any compensation. In general, for $0\leq g\leq\pi$, we therefore need to consider a compensation $e^{ih(g,\xi,\phi)\hat \sigma^x_{\rm qb}}$ that rotates the obtained ion state to the target state corresponding to $\xi=0$. It depends on the unknown $\phi$ and reads
\begin{align}
    h(g,\xi,\phi) = \textrm{atan}\frac{\textrm{sin}\tfrac{g}{2}\,\textrm{sin}\tfrac{\xi}{2}}{\textrm{cos}\tfrac{\xi}{2}\,(\textrm{cos}\tfrac{g}{2}\,\textrm{sin}\phi+1) - \textrm{cos}\tfrac{g}{2}\,\textrm{sin}\frac{\xi}{2}\,\textrm{cos}\phi}.
\end{align}
Only for the aforementioned cases $h(0,\xi,\phi)=0$ and $h(\pi,\xi,\phi)=\frac{\xi}{2}$ does this dependence vanish.

The standard assumption in this type of phase estimation task is to already have narrow prior information on $\phi$. With this assumption, we also expect that the compensation will work reasonably well when inserting the current estimate $\hat{\phi}$ of $\phi$, i.e.\ $h(g,\xi,\phi) \approx h(g,\xi,\hat{\phi})$.

With that assumption and with $0\leq\phi<\pi$, the ion state after $n$ repetitions (approximately) reads
\begin{align}
    \ket{\Psi_\text{fin}}_{\rm qb} &= \tfrac{1}{\sqrt{2(1+\cos(g/2)\sin(\phi))^n}}\big((\cos\tfrac{\phi}{2}+e^{\frac{i g}{2}} \sin \tfrac{\phi}{2})^n\ket{+}_{\rm qb}\nonumber\\
    &\ \ \ +(\cos\tfrac{\phi}{2}+e^{\frac{-i g}{2}} \sin \tfrac{\phi}{2})^n\ket{-}_{\rm qb}\big).
\end{align}
Ultimately, measuring this state in the computational basis yields
\begin{align}
    p_0(\phi) = \cos^2\!\big[n\, \textrm{acot}\!\big(\!\csc(\tfrac{g}{2})\,\textrm{cot}\big(\tfrac{\phi}{2}\big)+\cot (\tfrac{g}{2})\big)\big]. \label{eq:p_0}
\end{align}
Using the definition in Eq.~\eqref{eq: formula_FI}, this leads to a Fisher information of
\begin{align}
    F(\phi)=\frac{n^2\sin^2(\tfrac{g}{2})}{\big(1+\textrm{cos}(\tfrac{g}{2})\sin(\phi)\big)^2}, \label{eq:Fisher}
\end{align}
which recovers $F(\phi)=n^2$ for $g=\pi$ as well as $F(\phi)=0$ for $g=0$. In addition, for the standard task of estimating $\phi$ close to $0$, this simplifies to an attenuation of the Fisher information by a factor $\sin^2(\tfrac{g}{2})$. We further see that for all $\phi$, a metrological advantage $F(\phi)>n$ is achieved as soon as $n>(1+\textrm{cos}(\tfrac{g}{2})\sin(\phi))^2/\sin^2(\tfrac{g}{2})$.

The regime $g<\pi$ also comes with advantages, especially in the context of electron loss, which induces mixedness of the ion state. In order to see this, let the ion be prepared in the initial state
\begin{align}
    \rho_{\rm qb} = \tfrac{1}{2}\big(&\ketbra{+}_{\rm qb}+\ketbra{-}_{\rm qb}\nonumber\\
    &+s e^{i\beta_0}\ketbra{+}{-}_{\rm qb}+s e^{-i\beta_0}\ketbra{-}{+}_{\rm qb}\big). \label{eq:ion_initial}
\end{align}
After the interaction, we can trace out the electron to model its loss. The resulting density matrix of the ion has the exact same parametrization as (\ref{eq:ion_initial}), but the off-diagonal entries are updated according to ${s e^{\pm i\beta_0}\mapsto se^{\pm i\beta_0}(1+e^{\pm i g})/2}$. With another correction $\hat U=e^{-\frac{i}{4}g\hat \sigma^x}$, the off-diagonal entries get an additional factor $e^{\mp \frac{i}{2} g}$. This simplifies the mapping to ${s e^{\pm i\beta_0}\mapsto s e^{\pm i\beta_0}\cos(\tfrac{g}{2})}$. The phase information is thus preserved, but the amplitude of these terms decreases. This aligns with the intuition that for no entanglement ($g=0$), losing an electron has no effect, and for maximal entanglement ($g=\pi$), we need to restart after even a single electron loss. In contrast, for intermediate $g$, we have the advantage that we can tolerate the occasional loss of an electron.

Lastly, let us analyze the effect of electron loss on the Fisher information for $g\leq\pi$. Based on the above, we can rewrite the ion state as
\begin{align}
    \rho_{\rm qb}=(1-c)\tfrac{1}{2}\mathbbm{1} + c\,\rho_{\rm qb}^*,
\end{align}
where $\rho_{\rm qb}^*$ would be the ion state without loss, i.e.\ Eq.~\eqref{eq:ion_initial} with $s=1$, and ${c=\cos^m\tfrac{g}{2}}$ is the coefficient corresponding to the loss of $m$ electrons. Note that the order of lost and detected electrons does not matter. Next, we use the fact that the contribution of $\rho_{\rm qb}^*$ returns the original $p_0(\phi)$, Eq.~\eqref{eq:p_0}, while the term proportional to the maximally mixed state will always give $p_0=\tfrac{1}{2}$. From this, we derive that the loss of electrons can be accounted for by a multiplicative factor to the Fisher information without loss Eq.~\eqref{eq:Fisher}:
\begin{align}
    \tilde{F}(\phi)&=\frac{\left[\partial_{\phi}((1-c)\tfrac{1}{2}+c\, p_0(\phi))\right]^2}{((1-c)\tfrac{1}{2}+c\,p_0(\phi))\,(1-(1-c)\tfrac{1}{2}-c\,p_0(\phi))}\nonumber\\
    &=\frac{\left(\partial_{\phi} p_0(\phi)\right)^2}{p_0(\phi)\,(1-p_0(\phi))}\frac{4c^2\,p_0(\phi)\,(1-p_0(\phi))}{1-c^2\,(1-2p_0(\phi))^2}\nonumber\\
    &=F(\phi)\frac{4c^2\,p_0(\phi)\,(1-p_0(\phi))}{1-c^2\,(1-2p_0(\phi))^2}.
\end{align}

\end{document}